\documentclass[
reprint,
superscriptaddress,
showpacs,
amsmath,
amssymb,
aps,
pra,
longbibliography
]{revtex4-1}

\usepackage{graphicx}% Include figure files
\usepackage{dcolumn}% Align table columns on decimal point
\usepackage{bm}% bold math
\usepackage{color}% Allows color text
\usepackage{ulem}% allows strike-out using sout
\usepackage[caption=false]{subfig}

\newif\ifcmnt
\cmnttrue

\ifcmnt
    \providecommand{\aucmnt}[1]{#1}
\else
    \providecommand{\aucmnt}[1]{}
\fi

\newcommand{\rhotrue}{\rho_{\text{true}}}

\begin{document}

% \preprint{APS/123-QED}

\title{Investigating Bias in Maximum Likelihood Quantum State Tomography}% Force line breaks with \\
\author{G. B. Silva}
\affiliation{Departamento de Engenharia de Teleinform\'atica, Universidade Federal do Cear\'a, Fortaleza, Cear\'a, Brazil}
\author{S. Glancy}
\affiliation{Applied and Computational Mathematics Division, National Institute of Standards and Technology, Boulder, Colorado, 80305, USA}
\author{H. M. Vasconcelos}
\email{hilma@ufc.br}
\affiliation{Departamento de Engenharia de Teleinform\'atica, Universidade Federal do Cear\'a, Fortaleza, Cear\'a, Brazil}


\date{\today}% It is always \today, today,

\begin{abstract}
  Maximum likelihood quantum state tomography yields estimators that
  are consistent, provided that the likelihood model is correct, but
  the maximum likelihood estimators may have bias for any finite data
  set. The bias of an estimator is the difference between the expected
  value of the estimate and the true value of the parameter being
  estimated. This paper investigates bias in the widely used maximum
  likelihood quantum state tomography. Our goal is to understand how
  the amount of bias depends on factors such as the purity of the true
  state, the number of measurements performed, and the number of
  different bases in which the system is measured. For that, we
  perform numerical experiments that simulate optical homodyne
  tomography under various conditions, perform tomography, and
  estimate bias in the purity of the estimated state.  We find that
  estimates of higher purity states exhibit considerable bias, such
  that the estimates have lower purity than the true states.
\end{abstract}

 % PACS, the Physics and Astronomy Classification Scheme.
\maketitle

%\tableofcontents

\section{Introduction}
\label{intro}
Quantum state tomography (QST) is the estimation of an unknown quantum
state from experimental measurements performed on a collection of
quantum systems all prepared in the same unknown state. QST is an
important procedure for quantum computation and
information~\cite{banaszek2013}, being used, for example, to learn
properties of states prepared in experiments and for the validation of
quantum gates in quantum process tomography.

In QST many identical copies of the system are prepared, each copy is
independently measured, and the results of these measurements are used
to estimate the system's quantum state $\rhotrue$. A
commonly used method to make the estimate is maximum likelihood
estimation, in which one finds the state
$\rho_{\text{ML}}$ with the maximum likelihood given the measurement
results~\cite{Hradil}. The estimation is an optimization problem
usually solved numerically using iterative algorithms, such as the
expectation-maximization based $R \rho R$~\cite{Rehacek2007} and
gradient ascent algorithms. This optimization problem becomes more
difficult as the dimension of $\rhotrue$ increases.

In this paper, we examine idealized simulated experiments with no
systematic experimental errors, meaning that they are correctly
described by the likelihood model.  We analyze only the properties of
the random measurement error and bias in the maximum likelihood
estimator. Properties of this estimator have also been examined in
\cite{Sugiyama2012, Braunstein1994, Helstrom1967}.  The difference
between the estimate's expected value and the true value of the
parameter being estimated is called ``bias''. Given a correct
likelihood model and an informationally complete set of measurements
\cite{Hradil}, maximum likelihood estimators are consistent and 
asymptotically unbiased~\cite{Shao1998}, but they are typically
biased for finite samples.  This bias is caused by nonlinearity of the
estimation procedure.

We could be tempted to avoid the bias problem by using linear
inversion estimators, which are unbiased.  However, common linear
inversion estimators do not confine their estimates to physical state
space\cite{Shang2014}, and linear estimators generally have larger
mean squared error than maximum likelihood estimators~\cite{Shao1998}.

In~\cite{Sugiyama2012}, the behavior of estimation errors in one-qubit
state tomography was analyzed numerically using distances between the
estimate and the true state. That analysis showed that for the
tomography of a single qubit, the constraint that density matrices be
positive semi-definite creates bias that increases as the length of
the Bloch vector, a measure of distance from the state to the boundary
of state space, approaches 1.  However, that bias can be reduced if
measurement operators are aligned with the Bloch vector.
In~\cite{Schwemmer2015}, Monte Carlo simulations were used to study
quantum state tomography of a few qubits measured in the bases of the
Pauli operators. \cite{Schwemmer2015} showed that reconstruction
schemes based on maximum likelihood and least squares both suffer from
bias. The fidelity was systematically underestimated while the
entanglement was overestimated.

In this paper we are interested in the tomography of continuous
variable systems, whose bias has not yet been systematically
investigated.  We use numerical experiments to simulate optical
homodyne tomography data under various conditions and perform maximum
likelihood tomography on that data.  Because \cite{Sugiyama2012}
showed a relationship between bias and the length of a qubit's Bloch
vector, we extend those results to higher dimensional systems by
examining bias's dependence on purity, another measure of distance of
a density matrix from the boundary of state space.  We also
investigate how the amount of bias depends on factors such the number
of measurements performed, the number of different bases in which the
system is measured, and the dimension of the Hilbert space.  In
Section \ref{MLE} we review maximum likelihood estimation.  In Section
\ref{numerical-experiments} we describe our numerical experiments and
present our results.  In Section \ref{conclusion} we discuss the
interpretation of our results and make some concluding remarks.

\section{Maximum likelihood estimation}
\label{MLE}
Continuous variable systems, such as a harmonic oscillator or a mode
of light, live in infinite dimensional Hilbert spaces.  Tomography
cannot estimate the infinitely many parameters required to represent
states in infinite dimensional spaces, so the standard approach in
this case is to limit the number of unknown parameters, by truncating
the Hilbert space at a maximum phonon or photon number
$n$.

Let us consider $N$ quantum systems, each of them prepared in a state
described by a density matrix $\rhotrue$. Each copy $i$ has
an observable labeled by $\theta_i$ measured with result $x_i$,
for $i = 1, ..., N$. In each measurement the observable is chosen by
setting the phase $\theta_i$ of a local oscillator (a reference system
prepared in a high amplitude coherent state). The outcome of the
$i$-th measurement is described by a positive-operator-valued measure
(POVM) element $\Pi (x_i|\theta_i) = \Pi_i$. The likelihood of a
candidate density matrix $\rho$ given the data set \{$(\theta_{i},
x_i): i = 1, ..., N$\} is given by
\begin{eqnarray}
\mathcal{L} (\rho)= \prod_{i=1}^{N} \mathrm{Tr} (\Pi_i \rho),
\end{eqnarray}
where $\mathrm{Tr}(\rho \Pi_i)$ is the probability, according to
$\rho$, to obtain outcome $x_i$ when measuring with phase $\theta_i$.

The goal of maximum likelihood QST is to find the density matrix
$\rho_{\text{ML}}$ that maximizes the likelihood.  In practice
one usually maximizes the logarithm of the likelihood (the
``log-likelihood''):
\begin{eqnarray}
L (\rho) = \ln \mathcal{L} (\rho)= \sum_{i=1}^{N} \ln [\mathrm{Tr} (\Pi_i \rho)],
\end{eqnarray} 
which is maximized at the same density matrix as the likelihood. The
log-likelihood function is concave, giving us a well-behaved
optimization problem, such that the convergence to the unique solution
will be achieved by most iterative optimization methods.

Our algorithm for likelihood maximization begins with several
iterations of the $R\rho R$ algorithm followed by iterations of a
regularized gradient ascent algorithm (RGA). After an initial period
of fast convergence, we have observed significant slow-down in the
$R\rho R$ algorithm after around $(n+1)^2/4$ iterations. To
alleviate this problem, after $(n+1)^{2}/4$ $R\rho R$ iterations, we
switch to the RGA.  Let $\rho^{(k)}$ be the density matrix found after
$k$ iterations, the first of which is provided by the last iteration
of $R \rho R$. In the RGA, $\rho^{(k+1)}$ is parametrized as
\begin{equation}
  \rho^{(k+1)}=\frac{\left(\sqrt{\rho^{(k)}}+A\right)\left(\sqrt{\rho^{(k)}}+A^{\dagger}\right)}{\mathrm{Tr}\left[\left(\sqrt{\rho^{(k)}}+A\right)\left(\sqrt{\rho^{(k)}}+A^{\dagger}\right)\right]},
\end{equation}
where $A$ may be any complex matrix of the same dimensions as $\rho$.
This construction ensures that $\rho^{(k+1)}$ is a physical density
matrix for any $A$.  To choose $A$, a quadratic approximation of the
log-likelihood is performed.  $A$ maximizes the quadratic
approximation of the log-likelihood subject to the constraint that
$\text{Tr}(AA^{\dagger})\leq u$, where $u$ is a positive number that
the algorithm adjusts to ensure that the log-likelihood increases with
each iteration.

All iterations halt when the stopping criterion of \cite{Glancy2012}
signals that the $L(\rho_{\text{ML}})-L(\rho^{(k)})\leq 0.2$, where
$L(\rho_{\text{ML}})$ is the maximum of the log-likelihood.  By
bounding the log-likelihood improvement that can be achieved with
further iterations, we ensure that the last iteration produces an
estimate that is ``close'' to $\rho_{\text{ML}}$, where that closeness
is statistically relevant \cite{Glancy2012}.

\section{Numerical experiments}
\label{numerical-experiments}
Our numerical experiments simulated single mode optical homodyne
measurements~\cite{Lvovsky2009} of a state created by sending a
squeezed vacuum state, with quadratures variances $s/2$ and $1/(2s)$,
through a lossy medium with transmissivity $t$.  These states are
Gaussian states with zero means, which can be parametrized by their
covariance matrices. Since we want to simulate states of different
purities, we will express purity as a function of squeezing and
transmissivity.

The covariance matrix of the state after the lossy medium is given by
\begin{eqnarray}
\Sigma = t \left(
\begin{array}{cc}
\frac{1}{2s} & 0 \\
0 & \frac{s}{2} 
\end{array} \right) + (1-t)
\frac{\mathbb{I}}{2}, 
\end{eqnarray}
where $\mathbb{I}$ is the identity matrix. Purity is then given by~\cite{Paris2003}
\begin{align}
p(s,t) & = \frac{1}{2\sqrt{\mathrm{Det}(\Sigma)}} \nonumber \\
& = \frac{1}{2\sqrt{\left( \frac{1}{2} - \frac{1}{4s} - \frac{s}{4}\right) \left(t^2 - t\right) + \frac{1}{4}}}. 
\label{eq-p(s,t)} 
\end{align}

Our numerical experiments begin with the choosing of a desired purity
for the true state. We use Eq. (\ref{eq-p(s,t)}) to obtain an $(s,t)$
pair that produces the desired purity.  The choice of $(s,t)$ is not
unique, so we use two strategies, described below, that give states
that are close to the vacuum and highly squeezed states.  We
represent the pure squeezed state with squeezing $s$ as
$\rho_{\text{pure}}$, a density matrix in the photon number basis,
truncated at $n$ photons.  We then simulate passage of the pure
squeezed state through a medium with transmissivity $t$ by a quantum
operation that is equivalent to appending an ancillary mode in the
vacuum state, acting on the two modes with the beam splitter
transformation, and tracing-out the ancillary mode.  This quantum
operation is expressed with the set of Kraus operators
$\{E_i(t)|i=1\ldots n\}$, and transforms $\rho_{\text{pure}}$ into
$\rhotrue = \sum_{i=1}^n E_i(t)\rho_{\text{pure}}E_i(t)^\dagger$.
This procedure gives us the density matrix of a state with the desired
purity, represented in an $n$ photon basis.

To compute the probability
$P(x|\theta) = \mathrm{Tr}(\rhotrue \Pi(x|\theta))$ to obtain homodyne
measurement result $x$ at phase $\theta$ from state $\rhotrue$, we
derive a representation of $\Pi(x|\theta)$ also in the $n$ photon
basis.  Let $|x\rangle$ be the $x$-quadrature eigenstate with
eigenvalue $x$ expressed in the photon number basis, and let
$U(\theta)$ be the phase evolution unitary operator.  For an ideal
homodyne measurement, we would compute the probability as
$\mathrm{Tr}[\rhotrue U(\theta)^\dagger|x\rangle\langle x|U(\theta)]$.
However, real homodyne detectors suffer from photon loss.  Because
this loss is part of the measurement device, we include it in the POVM
elements by expressing them as
$\Pi(x|\theta) = \sum_{i=1}^n E_i(\eta)^\dagger U(\theta)^\dagger
|x\rangle \langle x| U(\theta) E_i(\eta)$ \cite{Lvovsky2004}.  By
including the loss associated with the measurement device in the POVM
elements, we can estimate the state of the system before that loss
occurs.  For all of our numerical experiments, we use $\eta = 0.9$,
which is typical for state-of-the-art homodyne detectors.  To produce
random samples of homodyne measurement results, we use rejection
sampling~\cite{Kennedy1980} from the distribution given by
$P(x|\theta)$.

We use two different methods for choosing the phases at which the
homodyne measurements are performed.  In the first method, for each
quadrature measurement a random phase is chosen.  In the second method for a total of
$N$ measurements, measured at $m$ different phases, we divide the
upper-half-circle evenly among the $m$ phases between 0 and $\pi$ and
measure $N/m$ times at each phase.  Measuring the quadrature only once
for each of very large number of phases is natural for experimental
systems that slowly scan the phase while sampling quadratures.  For
other systems, it may be more convenient to fix the phase and
repeatedly measure the quadrature before changing the phase.  We
expect these two strategies to behave differently for three
reasons. (1) When measuring evenly spaced phases, we
obtain a histogram of quadrature measurements that allows us to
directly reconstruct the probability distribution of the quadrature at
each of the phases, but when measuring random phases, our
knowledge of the quadrature probability distribution at each phase is
quite poor, but we obtain samples at many more phases.  The statistics
of the two strategies are quite different, and one might expect
that estimates produce different
biases. (2)~\cite{Sugiyama2012} showed that bias is reduced if one
measures a qubit in the direction of its Bloch vector.  By increasing
the number of phases at which we measure the optical mode, we increase
the probability that we measure in a direction that points toward the
boundary of state space, which might reduce bias. (3) To obtain a
single maximum of the likelihood function, we require an
informationally complete set of measurement operators.  If the number
of phases is too small relative to the maximum number of photons $n$
in the Hilbert space, we do not have an informationally complete set
of measurement operators, and there will be a family of density
matrices, all of which maximize the likelihood. According to
  \cite{Leonhardt1997}, $n+1$ different phases are required to
  reconstruct a state that contains at most $n$ photons.  The
likelihood maximization algorithm identifies one of these density
matrices, and there may be systematic error introduced in the process.

To calculate the mean purity, we reconstruct each state 50 times, each
time obtaining the purity for the reconstructed state. We then
calculate the arithmetic mean of the 50 purities to obtain the ``mean
reconstructed purity''. Our estimate of the purity bias is the
difference between the mean reconstructed purity and the purity of the
true state. The uncertainty in each bias estimate (shown as error bars
in the figures) is the standard deviation of the mean of the
reconstructed purity. The relative sizes of the magnitude of the bias
(abs(bias)) and the standard deviation of the 50 purities
(std(purity)) are also of interest, so we make some statements about
them in the figure captions.

\subsection{Nearly vacuum state}
\begin{figure}
\includegraphics[width=0.5\textwidth]{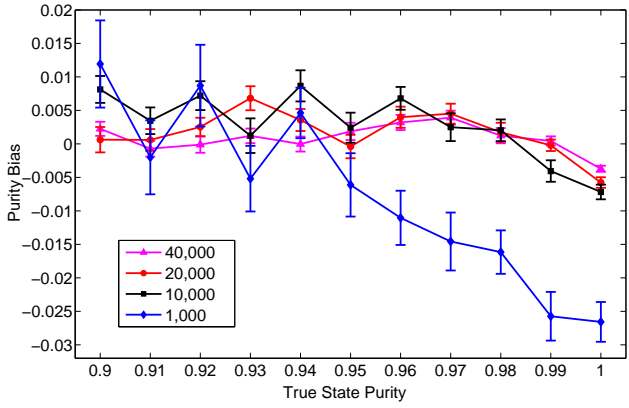}
\caption{\label{fig1}Purity bias as a function of true state purity
  for a nearly vacuum state and random phases. Number of
  measurements: 40,000, 20,000, 10,000, and 1,000. Maximum photon
  number: 10. In the 40,000, 20,000, and 10,000 measurement cases,
  abs(bias) is considerably smaller than std(purity), except they are
  approximately equal when purity=1 . In the 1,000 measurement case,
  abs(bias) and std(purity) are approximately equal at purity=0.99,
  and at purity=1 abs(bias) is larger than std(purity).}
\end{figure}

Let us start with tomography of a nearly vacuum state measured
  with randomly chosen phases. A state of a given purity with the
weakest squeezing is reached when $t=1/2$.  To find the necessary
squeezing, we solve Eq.~(\ref{eq-p(s,t)}) for $s$.  Fig.~\ref{fig1}
shows purity bias as a function of the true state purity for different
numbers of measurements. Analyzing the graphs, we can see that, when
fewer measurements are made, the highest purity states' mean
reconstructed purity is significantly lower than the true state
purity. This is clear evidence of bias in the tomography algorithm. We
can also see that as we increase the number of measurements, the
variance and bias of the purity estimates decreases, becoming
approximately unbiased asymptotically, as expected.

\begin{figure}
\begin{tabular}{c}
\subfloat[]{\includegraphics[width=0.5\textwidth]{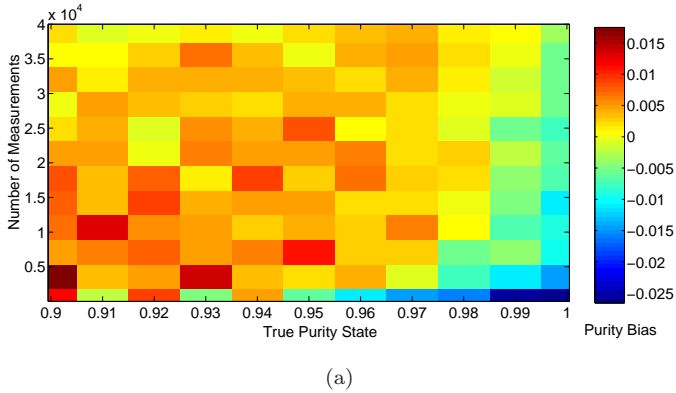}} \\
\subfloat[]{\includegraphics[width=0.5\textwidth]{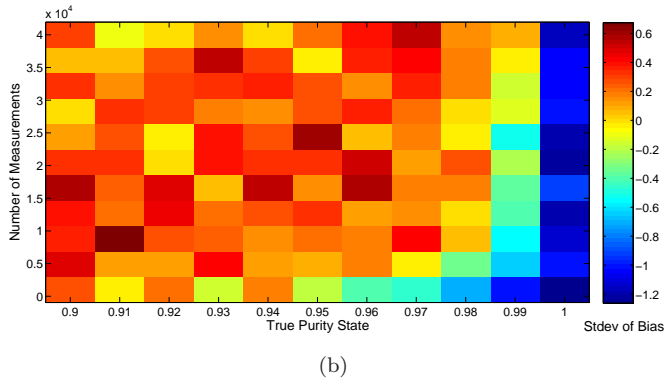}} 
\end{tabular}
\caption{Color map of the (a) purity bias and (b) purity bias
    measured in number of standard deviations of the purity estimates:
    (purity bias)/stdev(purity) as a functions of the number of
  measurements and the true state purity for a nearly vacuum state
  measured with random phases. Maximum photon number = 10.}
\label{fig-colormaps}
\end{figure}

Fig.~\ref{fig-colormaps} shows color maps of the purity bias as a function of the number of
measurements and the true state purity. These give us a qualitative
description of purity bias for a larger parameter space than shown in
Fig.~\ref{fig1}.

\begin{figure}
\includegraphics[width=0.5\textwidth]{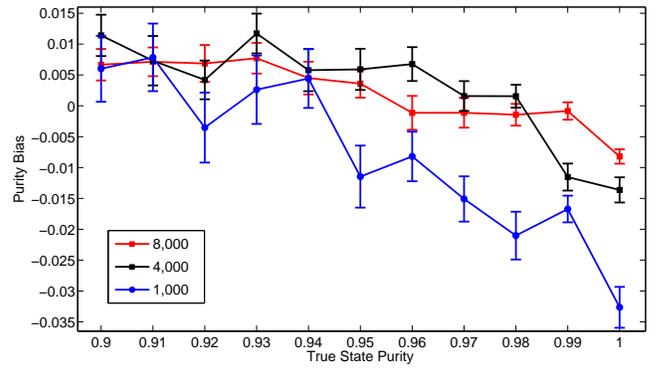}
\caption{\label{fig3}Purity bias as a function of true state purity
  for a nearly vacuum state and six evenly spaced phases. Number
  of measurements: 8,000; 4,000; and 1,000.  For 1,000
    measurements abs(bias) is larger than std(purity) for true state
    purities 0.99 and 1.00. For all other cases shown abs(bias) is
    less than std(purity), though several points are very close.}
\end{figure}

\begin{figure}
\includegraphics[width=0.5\textwidth]{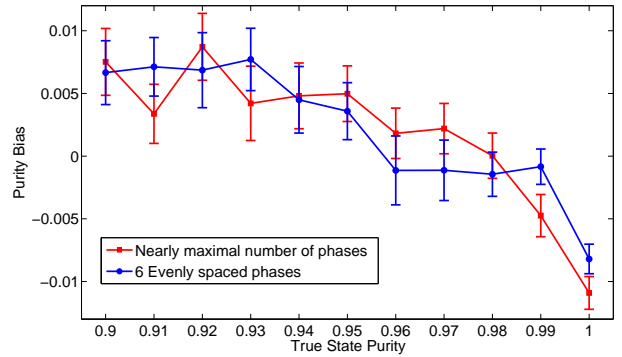}
\caption{\label{fig4} Purity bias as a function of true state purity
  for a nearly vacuum state and both phase choosing methods: random
  phases and six evenly spaced phases. Number of measurements:
  8,000. Maximum photon number: 10.}
\end{figure}

In Fig.~\ref{fig3}, we use a nearly vacuum state, but rather than
  randomly choosing a phase for each measurement, we used only six
  evenly spaced phases. 
In Fig.~\ref{fig4}, we plot the both 8,000 measurement case when using
random phases and the 8,000 measurement case with random phases (from
Fig.~\ref{fig3}).  Although only six phases
  is not informationally complete for the 10 photon Hilbert space in
  which we are performing the reconstruction, we see little effect on
  the purity bias.  This is likely because the number of photons in
  the state is much smaller than 10.

\begin{figure}
\includegraphics[width=0.5\textwidth]{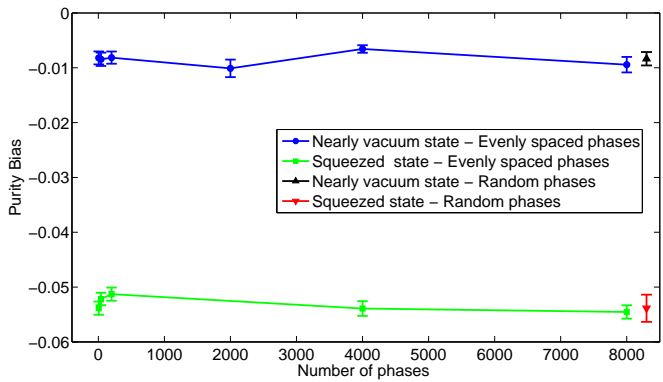}
\caption{Purity bias for reconstruction of states whose purity is 1 as
  a function of the number of evenly spaced phases at which the state
  is measured.  Bias for the vacuum state is shown with blue circles,
  and bias for a squeezed state whose squeezed quadrature variance is
  1/2 of the vacuum variance is shown with green squares.  Number of
  measurements = 8000.  Maximum photon number = 10.  The graph also
  shows the bias obtained when measuring 8000 random phases with the
  points just to the right of the 8000 evenly spaced phases point.
  Unlike other graphs in this paper, which use 50 simulated
  experiments to estimate bias, this graph uses 200 simulated
  experiments.}
\label{fig-m}
\end{figure}

To further explore the relationship between the number of phases and
bias, in Fig.~\ref{fig-m} we show the behavior of purity bias as a
function of the chosen number of evenly spaced phases.  Increasing the
number of phases has very little effect on bias.

\begin{figure}
  \includegraphics[width=0.5\textwidth]{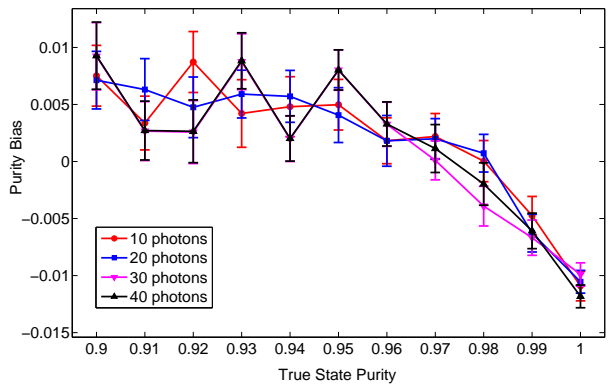}
  \caption{\label{fig5}Purity bias as a function of true state purity
    for a nearly vacuum state measured with random phases. Number of
    measurements: 8,000. Maximum photon number: 10, 20, 30 and 40. For
    all of these cases, we find abs(bias) larger than std(purity) only
    when the true state purity=1.}
\end{figure}

\begin{figure}
  \includegraphics[width=0.5\textwidth]{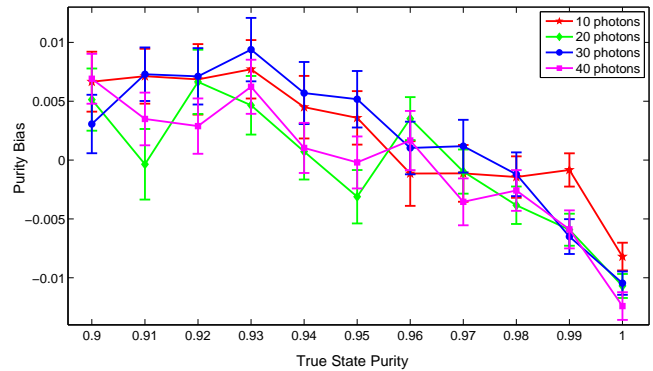}
  \caption{\label{fig6}Purity bias as a function of true state purity
    for a nearly vacuum state and 6 evenly spaced phases. Number of
    measurements: 8,000. Maximum photon number: 10, 20, 30 and 40.}
\end{figure}

So far, we have presented results of density matrices reconstructed in
a 10 photon Hilbert space. We now argue that 10 photons is sufficient
to represent the nearly vacuum states that we have analyzed. A state
will be well represented in a truncated Hilbert space of $n$ photons,
avoiding errors in the tomography, when the sum of probabilities of
having $n$ photons in this state is close to $1$. For the nearly
vacuum states, the number of photons in the state increases with
decreasing purity.  The nearly vacuum state with purity of 0.9 (the
lowest that we report) has a probability of $1.15 \times 10^{-5}$ to
contain more than 10 photons, so a 10 photon Hilbert space should
faithfully represent all of the nearly vacuum states. In
Fig.~\ref{fig5} we show purity bias of density matrices reconstructed
in 10 - 40 photon Hilbert spaces.  We see that the use of larger
Hilbert spaces has no effect on the purity bias.  Fig. \ref{fig5} also
shows that when randomly choosing a phase for each of 8,000
measurements, more than sufficient phase information is gained to
estimate the state even in a 40 photon Hilbert space.  We explore the
effect of measuring with an insufficient number of phases in
Fig.~\ref{fig6}, where we use only six evenly spaced phases to measure
the nearly vacuum state.  We reconstruct the density matrix in 10,
  20, 30, and 40 photon Hilbert spaces, and we see that the results
  are very similar to those obtained when measuring at random phases.

\subsection{Highly squeezed states}

To test the robustness of some of our claims, we now change the
measured states from nearly vacuum states to highly squeezed
states. Each state is created by sending a pure squeezed vacuum state,
whose squeezed quadrature has variance 1/4 of
vacuum variance, through a lossy medium with one of the following
transmissivities: $t = [0.5, 0.8, 0.9, 0.95, 0.99, 1]$. For each pair
$(s,t)$, the purity is calculated using Eq.~(\ref{eq-p(s,t)}). The
highly squeezed states contain more photons, so we truncate the
Hilbert space at 20 photons.  The highly squeezed state with the most
photons has purity of 1 (and $t=1$).  That state's probability to
contain more than 20 photons is $2.7 \times 10^{-6}$. The
results for bias of highly squeezed states are similar to those of the
nearly vacuum states.

\begin{figure}
\includegraphics[width=0.5\textwidth]{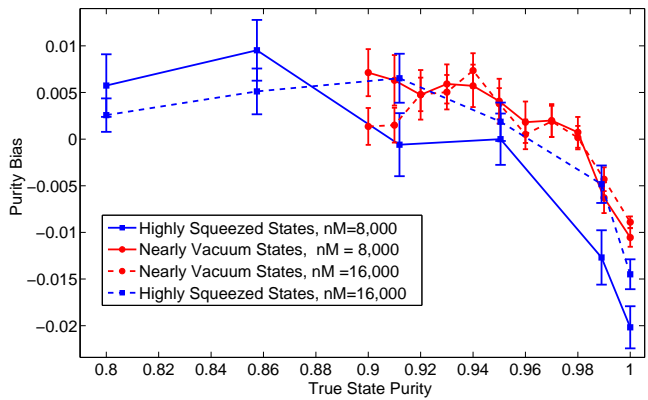}
\caption{\label{fig7}Purity bias as a function of true state purity
  for nearly maximal number of phases and both nearly vacuum and
  highly squeezed states.  A random phase is chosen for each
  measurement. Number of measurements: 16,000 and 8,000. Maximum
  photon number = 20.  For reconstructing nearly vacuum states from
  8,000 or 16,000 measurements abs(bias) is greater than std(purity)
  except when purity=1.  The same is true for reconstructing highly
  squeezed states from 8,000 measurements.  For reconstructing highly
  squeezed states from 16,000 measurements, abs(bias) is greater than
  std(purity) in all cases shown, except they are approximately equal
  when true state purity=0.95.}
\end{figure}

In Fig.~\ref{fig7} we compare the estimated purities when measuring
nearly vacuum and highly squeezed states, finding similar behavior in
the two cases.  It appears that bias is slightly higher for the highly
squeezed states. Bias is clearly not a function of the true state's
purity alone, but depends on other features of the true state.  This
dependence is not well understood.  As more measurements are taken the
biases decrease, and the gap between the bias of highly squeezed and
nearly vacuum states decreases.

We also show the dependence of bias on the number of evenly spaced
  phases used to measure a squeezed state in Fig.~\ref{fig-m}, seeing
  little relationship between bias and the number of phases, except
  there seems to be a slight increase in abs(bias) when very few
  phases are used.

\section{Conclusion}
\label{conclusion}

We have used idealized numerical experiments to generate simulated
data under various conditions, performed tomography, and estimated
bias in the purity of the results. The mean reconstructed
purities of the highest purity states are significantly lower than the
corresponding true state purities. This result shows clear evidence of
bias in the tomography algorithm, even when the likelihood model is
correct.  In our simulations we did not see a strong relationship
  between purity bias and the number of phases used to measure the
  state, though it is possible that such a relationship for some
  combination of states and measurement conditions. The abs(bias)
appears to be slightly larger for tomography performed on highly
squeezed states than in nearly vacuum states.

In this work we have focused on the influence of the true state's
purity on bias of the estimated purity, finding that more pure true
states suffer from more bias toward lower purity states.  This agrees
with the results of \cite{Schwemmer2015}. However, whether this is a
general property for all states is an open question.  More numerical
experiments on a greater diversity of states and using different
measurement schemes would be informative as would exploration of bias
in other parameters.

In many of our numerical experiments, we find that the bias in purity
is significant compared to the standard deviation of the estimates of
purity.  This is particularly problematic if tools like the bootstrap
are used to assign uncertainties in quantum state tomography.  If a
non-parametric bootstrap is used, every bootstrapped estimate will be
similarly biased.  If a parametric bootstrap is used, the original
estimate is biased once and the bootstrapped estimates will be biased
a second time.  Bias correction methods exist for parametric
bootstrap, but they require the bias to be consistent for different
states \cite{Efron1993}.  This might be a reasonable approximation,
but we have seen that it is not strictly true. Because of the problems
caused by bias, it maybe helpful to use confidence intervals such as
those described in \cite{Blume-Kohout2012, Christandl2012, Faist2015}
to assign uncertainties to estimated parameters.  Unfortunately those
methods produce confidence regions that are significantly larger (and
more conservative) than those produced by bootstrap methods commonly
used for quantum state tomography.

\begin{acknowledgments}
We thank Kevin Coakley, Adam Keith, and Emanuel Knill for helpful
comments on the manuscript.  H. M. Vasconcelos thanks the Instituto
Nacional de Ci\^encia e Tecnologia de Informa\c c\~ao Qu\^antica
(INCT-IQ). G. B. Silva thanks Coordena\c c\~ao de Aperfei\c coamento
de Pessoal de N\'ivel Superior (CAPES) for financial support. This
work includes contributions of the National Institute of Standards and
Technology, which are not subject to U.S. copyright.
\end{acknowledgments}

% Scott: My LaTeX does not know about spphys, but it is not necessary
% to specify a bibliography style.  revtex should authomatically use
% the correct style based on the documentclass.
\bibliography{bias}   % name your BibTeX data base

% Non-BibTeX users please use%\begin{thebibliography}{}

%\end{thebibliography}

\end{document}